\documentclass[12pt]{article}
\usepackage{graphicx}

\begin{document}

\begin{flushright}
CERN-PH-TH/2006-134
\end{flushright}
\vfill

\begin{center}
{\Large The spectrum of lattice QCD \\ with staggered fermions at strong
coupling} 
\vspace{1.0cm}

Philippe de Forcrand$^{a,b}$\footnote{e-mail:address:
forcrand@phys.ethz.ch} and Seyong Kim$^{c}$\footnote{e-mail address:
skim@sejong.ac.kr}

\vspace{1.0cm}

{\it $^a$ Institut f\"ur Theoretische Physik, ETH, CH-8093 Z\"urich, Switzerland } \\
{\it $^b$ CERN, Physics Department, TH Unit, CH-1211 Geneva 23, Switzerland } \\
{\it $^c$ Department of Physics, Sejong University, Seoul 143-747, Korea }

\vspace{2.0cm}

\end{center}

Using 4 flavors of staggered fermions at infinite gauge coupling,
we compare various analytic results for the hadron spectrum with
exact Monte Carlo simulations. Agreement with Ref.~\cite{Martin_etal}
is very good, at the level of a few percent.

Our results give credence to a discrepancy between the baryon mass and
the critical chemical potential, for which baryons fill the lattice at 
zero temperature and infinite gauge coupling.
Independent determinations of the latter set it at about 30\% less
than the baryon mass. One possible explanation is that the
nuclear attraction becomes strong at infinite gauge coupling.

\vfill

\thispagestyle{empty}
\pagebreak

\section{Introduction}

The strong coupling limit of lattice QCD can provide valuable insight,
as Wilson showed by proving the area-law behaviour for the Wilson loop
in that regime~\cite{Wilson}. Analytic integration over the gauge link
variables becomes feasible in this limit. Within some reasonable
approximation (typically mean field), one can then calculate
analytically the hadron spectrum. The result turns out to be not too
different from the continuum QCD spectrum. This has been the subject
of numerous
works~\cite{Saclay,Kawamoto_etal,Martin_etal,Jolicoeur}. To the best
of our knowledge however, these analytic results involving different
approximation schemes have never been compared with the results of
numerical simulations in the infinite coupling limit
($\beta\equiv6/g^2 = 0$). Here, we perform such a comparison for the
case of 4 flavors of staggered fermions with SU(3) gauge group.

In addition, we are motivated by a puzzle in the phase diagram of QCD
as a function of temperature $T$ and chemical potential $\mu$. This
phase diagram also can be determined analytically when $\beta=0$, and
has been investigated by use of increasingly sophisticated
approximations over the
years~\cite{Damgaard_etal}-\cite{Kawamoto2005}. At zero temperature, a
first-order transition is predicted to take place, where the baryon
density jumps from zero to saturation (1 baryon per lattice site). The
corresponding critical quark chemical potential for massless quarks
has been calculated in the mean-field approximation, which yields $
\mu_c a \sim 0.66$~\cite{Damgaard_etal}.  This analytic prediction has
also been confirmed by numerical simulation of a gas of loops, the
monomer-dimer-polymer ensemble, whose partition function is identical
to that of lattice QCD with staggered fermions at
$\beta=0$~\cite{Karsch_Mutter}, and by the Glasgow reweighting
method~\cite{Barbour_etal}. The numerical simulation
of~\cite{Karsch_Mutter} gives $\mu_c a = 0.63(2)$ in the chiral limit
and $0.690(15)$ at $m_q a = 0.1$. The result from~\cite{Barbour_etal}
is $\mu_c a = 0.687(15)$ at $m_q a = 0.1$. At first sight, this
excellent agreement between mean-field and Monte Carlo is very
satisfying. On second thought, however, it is rather mysterious. One
would expect the transition to occur when ($3\mu$) exceeds the free
energy required to create an additional baryon, which is equal to its
mass at $T=0$. Now, every analytic prediction for the nucleon mass
$m_N$ gives a value $(m_N a)$ very close to 3 so that $(\mu_c a)$
should be very close to 1. The values reported above are
considerably smaller. Among the various explanations for this puzzle,
we want to check the validity of the analytic approximations:
is the nucleon mass really close to 3 in the strong coupling limit, 
or is it close to $3 \mu_c a$ instead?

In the next section (Sec.II), we review the analytic approximations
which have been proposed for the hadron spectrum. Then (Sec.III) we
present our simulation results and compare our numerical results to
the analytic predictions in Table I. Conclusions follow.

\section{Overview of analytic approaches}

In the strong coupling limit, the gauge field part of the action can
be dropped because its coefficient is $\beta = 0$, and the
gauge link variables become random $SU(N)$ group elements. Then, gauge
links give non-vanishing contributions only when a given link forms a
part of an $SU(N)$ singlet combination. Using this property, one can
construct either an effective action and derive the hadron spectrum from
it~\cite{Saclay,Kawamoto_etal}, or one can count the graphically relevant
diagrams and obtain from them the hadron spectrum~\cite{Martin_etal}.

In~\cite{Saclay,Kawamoto_etal}, the authors derive the effective action
for $SU(3)$ in terms of meson and baryon degrees of freedom by
integrating the gauge link variables out. Then, via systematic large-$d$
(the spacetime dimension) expansion of this effective action,
the chiral condensate and the correlation functions of the composite
mesonic and baryonic fields are obtained. The positions of poles in the
zero momentum projected correlators give meson ($M_M$) and baryon ($M_B$)
masses which can be summarized as follows:
\begin{eqnarray}
\frac{1}{N} \langle \overline{\psi}\psi \rangle &\approx&
\sqrt{\frac{2}{d}}(\overline{\lambda} - 2 \overline{m}) \\
\cosh M_M a &=& d (\overline{\lambda}^2 - 1) + 2 k + 1, \\
\sinh M_B a &=& \frac{1}{2} \overline{\lambda}^N (2d)^{\frac{N}{2}}
\end{eqnarray}
where $\overline{\lambda} = \overline{m} +
\sqrt{\overline{m}^2 + 1}$, $\overline{m}= ma/\sqrt{2d}$,
$m a$ is the dimensionless quark mass, 
and $k=0,1,2,3$ gives access to four meson channels, to which are
assigned the physical meaning of the $\pi$, $\rho$, $a_1$, $a_0/f_0$
respectively.
For small $m a \ll \sqrt{2d}$,
\begin{eqnarray}
 \langle \overline{\psi}\psi \rangle &\approx& 2.12 - 0.75 m a \\
(M_\pi a)^2 &\approx& 5.66 m a \\
 M_\rho a &\approx& 1.76 + m a \\
 M_{a_1} a &\approx& 2.29 + 0.58 m a \\
 M_{a_0/f_0} a &\approx& 2.63 + 0.41 m a \\
 M_B a &\approx& 3.12 + 1.06 m a 
\end{eqnarray}

Leading $1/d$ corrections were obtained in~\cite{Jolicoeur}, yielding for
$d=4$, $N=3$:
\begin{eqnarray}
 \langle \overline{\psi}\psi \rangle &\approx& 1.99 - 0.56 m a \\
(M_\pi a)^2 &\approx& 4.60 m a \\
 M_B a &\approx& 2.93 + 1.99 m a 
\end{eqnarray}
and the masses of the other 3 mesons are unchanged.
Leading corrections in $1/g^2$ were also obtained in Ref.~\cite{Jolicoeur}.
They always come in the combination $1/(g^2 N)$.

Perhaps this motivated the authors of~\cite{Martin_etal} to try a
different approach. They developed a graphical method for summing
relevant, tree-like diagrams in the large-$N$ limit ($g^2 N
\rightarrow \infty$) at fixed $d$. They showed
\begin{eqnarray}
\frac{1}{N} \langle \overline{\psi}\psi \rangle &=&
\frac{d m_r - (2d-1) ma}{d^2 + (ma)^2} \\
\cosh M_M a &=& \frac{((2d-1)-(2d-2){m_r}^2 + {m_r}^4)}{2 {m_r}^2} + 2 k, \\
\sinh M_B a &=& \frac{1}{2} {m_r}^N 
\end{eqnarray}
where $m_r = m a + \sqrt{(m a)^2 + 2d-1}$ and $k=0,1,2,3$.
Again, for small $m a \ll \sqrt{2d - 1}$, one obtains
\begin{eqnarray}
(M_\pi a)^2 &\approx& 4.54 m a \\
M_\rho a &\approx& 1.76 + 0.81 m a \\
M_{a_1} a &\approx& 2.63 + 0.32 ma \\
M_{a_0/f_0} a &\approx& 2.29 + 0.46 ma \\ 
M_B a &\approx& 2.91 + 1.14 m a
\end{eqnarray}

The three sets of predictions are collected in Table I below, where
they are denoted respectively ``M.F.''(mean field)
~\cite{Saclay,Kawamoto_etal}, ``$1/d$-corr.''~\cite{Jolicoeur}
and ``$d=4$, large-$N$''~\cite{Martin_etal}. The last two columns
show our Monte Carlo results, presented in the next section,
for the quenched and the full QCD theories. 

\begin{table*}[htb]
\caption{Comparison of hadron properties.}
\vspace*{0.4cm}
\label{spectrum}
\begin{tabular}{|c|l|lllll|}
\hline
& $m_q a $ & M.F. & $1/d$-corr. & $d=4$ &MC results&MC results\\
&        & \cite{Saclay,Kawamoto_etal} & \cite{Jolicoeur} & large-$N$ \cite{Martin_etal} & quenched & $N_f=4$ \\
\hline
average         & 0.1   &  --  &  --  &  --        & 0.00001(4) &0.00603(2) \\
plaquette       & 0.05  &  --  &  --  &  --        &  same      &0.00656(3) \\
                & 0.025 &  --  &  --  &  --        &  same      & 0.00682(3)\\
\hline
                                       & 0.1 & 2.0463 & 1.9325 &
1.9281 &  1.9295(5) &1.9172(2) \\ 
$\langle \overline{\psi} \psi \rangle$ & 0.05 &2.0838 & 1.9606 &
1.9562 &  1.9574(7) &1.9413(3) \\
                                       & 0.025 &2.1026 & 1.9747 &
1.9703 &  1.9710(11)&1.9520(6)\\
\hline
        & 0.1   & 0.7521 & 0.6780 & 0.6735 & 0.6776(1) & 0.6780(1)\\
$m_\pi$ & 0.05  & 0.5318 & 0.4794 & 0.4762 & 0.4778(1) & 0.4784(1)\\
        & 0.025 & 0.3761 & 0.3390 & 0.3367 & 0.3374(1) & 0.3379(1)\\
\hline
         & 0.1   & 1.863 & 1.863 & 1.841 & 1.847(2) & 1.831(2)\\
$m_\rho$ & 0.05  & 1.813 & 1.813 & 1.801 & 1.802(4) & 1.774(4)\\
         & 0.025 & 1.788 & 1.788 & 1.780 & 1.768(6) & 1.779(7)\\
\hline
          & 0.1   & 2.350 & 2.350 & 2.336 & 2.337(6)  & 2.276(6)\\
$m_{a_1}$ & 0.05  & 2.321 & 2.321 & 2.313 & 2.317(9)  & 2.274(10)\\
          & 0.025 & 2.307 & 2.307 & 2.302 & 2.325(18) & 2.300(19)\\
\hline
               & 0.1   & 2.675 & 2.675 & 2.662 & 2.637(16) & 2.511(23) \\
$m_{a_0/f_0}$  & 0.05  & 2.654 & 2.654 & 2.646 & 2.588(27) & 2.551(27) \\
               & 0.025 & 2.644 & 2.644 & 2.638 & 2.598(51) & 2.636(47) \\
\hline
      & 0.1   & 3.225 & 3.129 & 3.024 & 2.961(3) & 2.931(3)\\
$m_N$ & 0.05  & 3.172 & 3.030 & 2.967 & 2.890(4) & 2.863(5)\\
      & 0.025 & 3.146 & 2.980 & 2.939 & 2.883(8) & 2.831(10)\\
\hline
\end{tabular}
\end{table*}

\section{Numerical results}

We have simulated the partition function $Z = \int {\cal D}U~e^{-S}$,
with the staggered action $S$:
\begin{equation}
S = \sum_x [ \frac{1}{2} \sum_{\mu} \eta_\mu (x) \{
  \overline{\psi} (x) U_\mu (x) \psi (x+\hat{\mu}) - \overline{\psi}
  (x+\hat{\mu}) {U_\mu}^\dagger (x) \psi (x)\} + m a
  \overline{\psi} \psi 
  (x) ] \label{action} 
\end{equation}
with $\eta_\mu(x) = (-)^{x_1 + .. + x_{\mu - 1}}$,
for 3 values of the quark mass ($m a = 0.1, 0.05$ and $0.025$), on an
$8^3 \times 16$ lattice. In each case, 500 decorrelated
configurations on which we measured the quark condensate and the
hadron correlators have been accumulated. Since, with staggered fermions, a
given hadron correlator also contains contributions from its parity
partner, we perform joint fits of the two correlators of parity
partners~\cite{Kim_Sinclair}:
\begin{eqnarray}
 C_1(t) &=& A (e^{-m_1 t} + e^{-m_1 (T-t)}) + (-1)^t B (e^{-m_2 t} +
 e^{-m_2 (T-t)})\nonumber \\
 C_2(t) &=& A' (e^{-m_2 t} + e^{-m_2 (T-t)}) + (-1)^t B' (e^{-m_1 t} +
 e^{-m_1 (T-t)}). \label{fitform}
\end{eqnarray}
For example, the $\pi$ channel correlator is fitted simultaneously with
the scalar channel correlator, and the $\rho$ channel correlator with the
pseudovector channel correlator by use of the 6-parameter fitting form of
Eq.(\ref{fitform}).

\begin{figure*}[ht]
\vskip 2.0cm
\begin{center}
\includegraphics[width=5.6in]{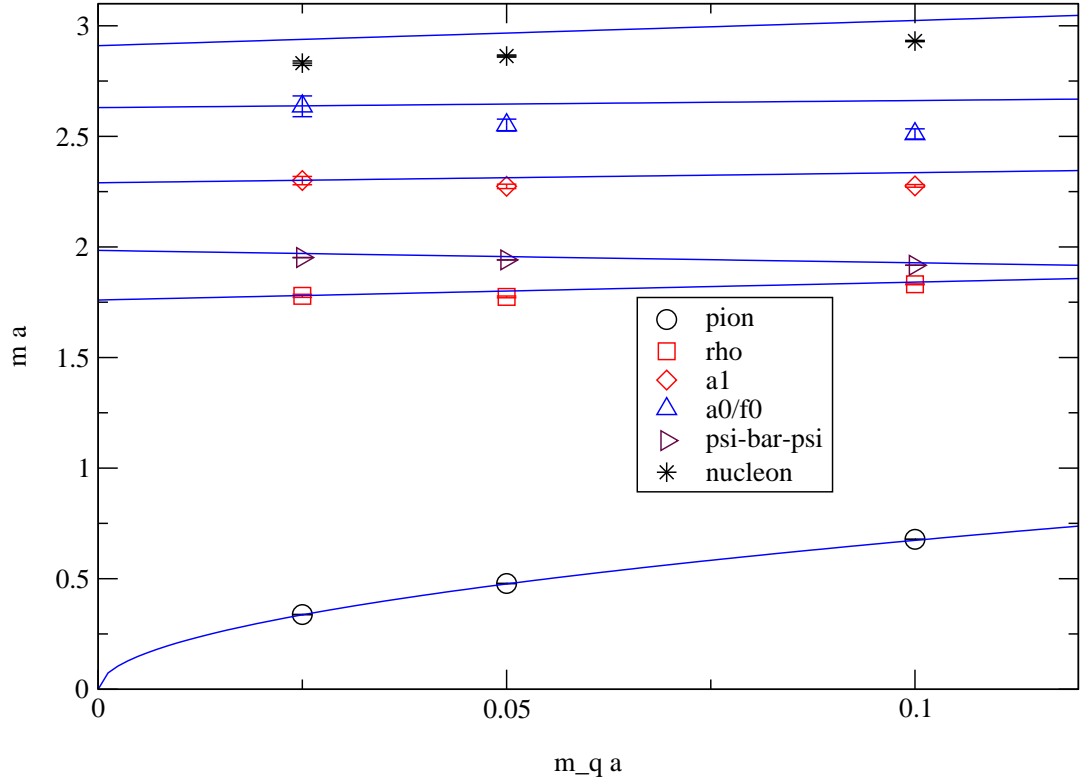}
\vskip 0.5cm
\caption{Hadron masses as a function of the quark mass $m_q a$.
The solid lines show the predictions of Ref.~\cite{Martin_etal}. }
\label{hadronmass}
\end{center}
\setcounter{figure}{1}
\end{figure*}

A few words on our simulation algorithm may be useful. The 4-flavor
theory can be simulated with standard Hybrid Monte
Carlo. Nevertheless, we used RHMC~\cite{RHMC1}, with two sets of
pseudo-fermion fields each yielding $\sqrt{\det(M)}$ after
integration. This decomposition allowed us to use a larger
stepsize~\cite{RHMC2}. More importantly, the stopping criterion of our
solver was set to the very loose value $1.0\times 10^{-2}$, while
maintaining an acceptance $\sim 65$\%. This great advantage of RHMC,
recognized in~\cite{RHMC2}, provides an important gain in efficiency.

\begin{figure*}[ht]
\vskip 2.0cm
\begin{center}
\includegraphics[width=5.6in]{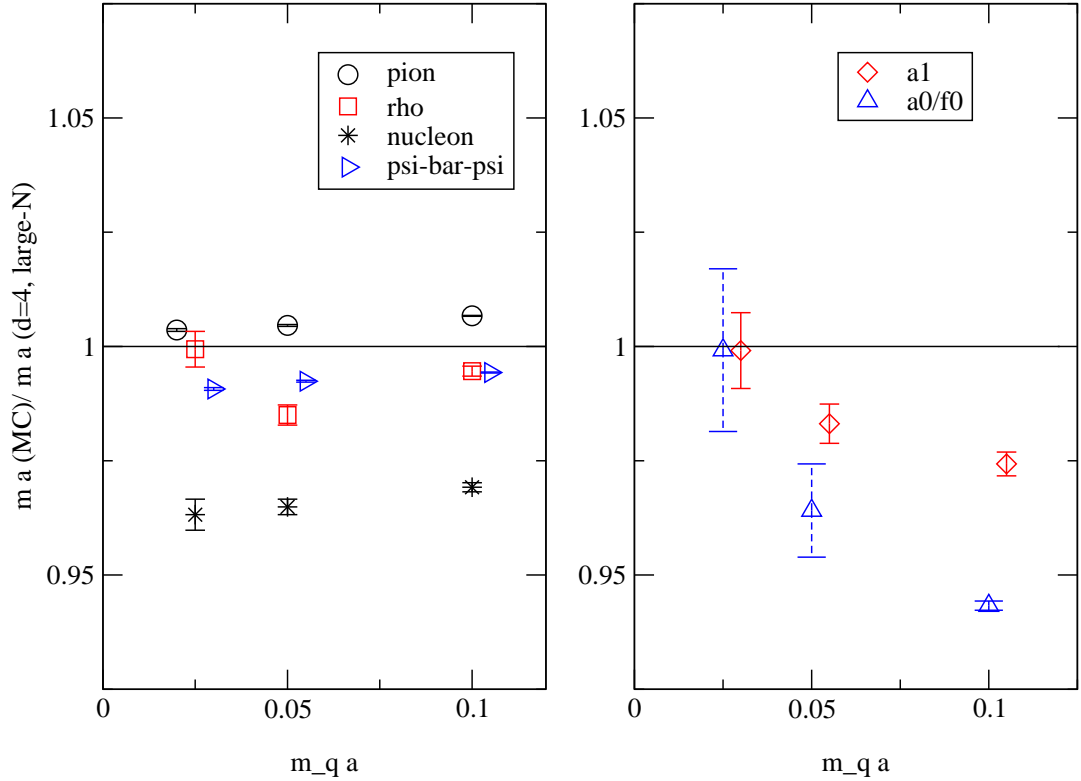}
\vskip 0.5cm
\caption{The ratio of our Monte Carlo data over the analytic result  
of Ref.~\cite{Martin_etal}, as a function of the quark mass $m_q
a$. Some channels are horizontally shifted to avoid overlapping data points.}
\label{massratio}
\end{center}
\end{figure*}

Our spatial volume is reasonably large compared to the pion mass
($m_\pi L = 2.6 $--$5.3$), and is very large compared to the nucleon
mass ($L \sim 5.5$ fm). Thus we expect negligible
finite-size effects on all observables except perhaps the pion mass,
and analytic predictions obtained on an infinite lattice
~\cite{Saclay,Kawamoto_etal,Martin_etal,Jolicoeur} can be directly compared
with our numerical results. 
Our $m_\pi/m_\rho$ mass ratio ranges from 0.3703(5) down to 0.1899(7),
compared to the physical value 0.179. So we are studying the light-quark
regime. Note, however, that $m_N/m_\rho$ is about 1.6, close to the
non-relativistic value $3/2$, and hardly changes with the quark mass.
So the spectrum of strong coupling QCD is only crudely similar to 
that of continuum QCD.

Our results are collected in the rightmost two columns of Table I,
for the quenched ($N_f=0$) and the full ($N_f=4$) theory.
Ref.~\cite{Martin_etal} counts all quark ``tree graphs'' which enclose
zero area and is very similar to the quenched approximation. This
motivated us to include quenched simulation results for comparison. 
Indeed, the tree graph resummation provides an excellent approximation 
to the quenched results\footnote{Compared to the quenched theory, mesonic 
tree graphs do not include the contribution of baryon loops, but these are 
suppressed by the heavy baryon mass. Baryonic tree graphs do not include 
the contribution of meson loops, which is probably the reason why the nucleon 
mass prediction is slightly wrong.}. 
Differences (of about 2\%) are visible in the nucleon mass only.
In turn, the quenched and full QCD results are very close to each other,
with the most difference (at the 5\% level) in the $a_0/f_0$ channel.

The agreement between the full QCD results and the mean-field
calculations~\cite{Saclay,Kawamoto_etal} is reasonable, and improves
with $1/d$ corrections~\cite{Jolicoeur}. It is best in general with
the large-$N$ approach of \cite{Martin_etal}.

\begin{figure*}[ht]
\vskip 1.0cm
\begin{center}
\includegraphics[width=5.6in]{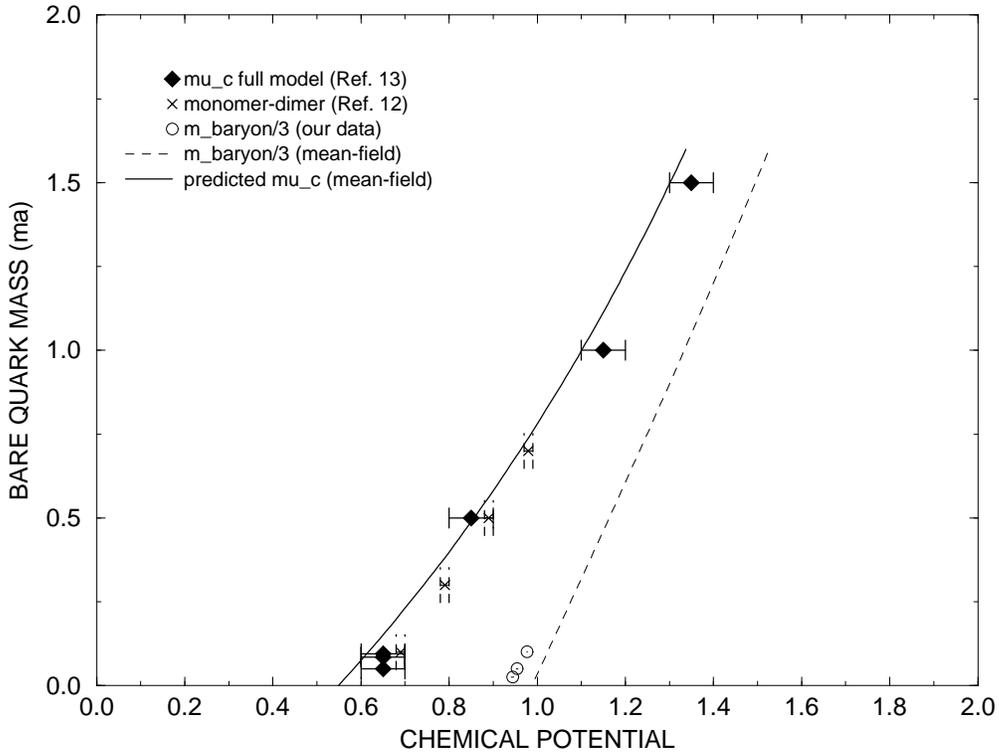}
\vskip 0.5cm
\caption{$m_{\rm baryon} a/3$ as a function of the quark mass $m_q a$ together
  with $\mu_c a$ (our data is added to the Figure from
  Ref.~\cite{Barbour_etal})}
\label{muc}
\end{center}
\end{figure*}

To illustrate the level of agreement with the predictions
of~\cite{Martin_etal}, we show in Fig.~1 the masses we measured
together with the analytic dependence of~\cite{Martin_etal} as a function
of the quark mass. In Fig.~2,
we show the ratio of our measured values over those predicted 
by~\cite{Martin_etal}. The agreement is better than within 1\% for the
chiral condensate $\langle \overline{\psi}\psi \rangle$ and the pion
mass. The ratio for the rho mass varies non-monotonically, which is 
possibly caused by the crossing of the threshold for $\rho \to \pi\pi$ decay.
For the nucleon, the Monte Carlo data is within 5\%
of~\cite{Martin_etal}. Our determination of the scalar mass
($a_0/f_0$) is too poor to provide a real check. Interestingly,
the lattice result agrees with the analytic result $m_\pi < m_\rho < m_{a_1}
< m_{a_0/f_0}$, in contrast to the experimental ordering, $m_\pi < m_\rho
< m_{a_0/f_0} < m_{a_1}$.

In Figure 3, we compare our nucleon mass with old Monte Carlo data 
for the critical, zero temperature quark chemical potential in infinite 
coupling QCD~\cite{Karsch_Mutter, Barbour_etal}. 
The error bar for $1/3$ of our nucleon mass is smaller than the plot
symbol. There is a clear, large difference between the critical quark
chemical potential and $1/3$ of the nucleon mass.

\section{Conclusions}

On the whole, our study justifies the mean-field approximation used
in~\cite{Saclay, Kawamoto_etal}. It also provides a beautiful
confirmation of the approach of~\cite{Martin_etal}, which goes beyond
mean-field and becomes exact when $g{^2}N \to \infty$. It may seem
strange in fact that this $g{^2}N \to \infty$ approximation is so good
(at the percent level), since virtual quark loops are absent in this
approximation in contrast to the numerical simulation. The
explanation, we believe, lies in the small plaquette average
value. The dynamical quarks shift the plaquette away from the value
zero which it would take in the quenched case, but only slightly. This
shift increases very mildly for lighter quarks. The lightest quarks
which we considered, $m_q a = 0.025$, give a pion to rho mass ratio
near the physical one, but yield a plaquette value of $0.00682(3)$
only. This is much less than the naive estimate based on a $1/(m_q a)$ 
expansion~\cite{Hasenfratz_DeGrand}, which would predict an effective shift
$\Delta\beta \approx 0.8$ in the gauge action, leading to a plaquette
value ${\cal O}(0.1)$, about 15 times larger than measured.
Thus, the price to pay for creating virtual quark loops remains high, 
and their effect on the spectrum is small. 
For the same reason, the quenched approximation turns out to be closer 
to full QCD at strong coupling than at weak coupling: at strong coupling, 
the ordering effect of the fermion loops is ``drowned'' in the disorder 
of the gluons.

As shown in Table I, the nucleon mass is just a bit smaller than
3. In particular, for $m_q a = 0.1$ which is the quark mass used in
the finite density simulations of~\cite{Karsch_Mutter,Barbour_etal},
the nucleon mass is 2.931(3) and one would expect the $T=0$ transition
induced by a quark chemical potential to occur at $\mu_c = 2.931(3)/3 =
0.977(1)$. However, the measured value for
$\mu_c$~\cite{Karsch_Mutter} is $0.690(15)$
($0.687(15)$ in~\cite{Barbour_etal}). The discrepancy with the expected
value is about 30\%. In other words, the nucleon seems to have a free
energy about 300 MeV less than its mass. Could this be caused by a
strong attraction among nucleons~\cite{Kogut,Bilic}? The weakness of the
real-world nuclear interaction results from near-cancellation between
the attractive omega exchange and the repulsive sigma
exchange~\cite{nuclear}. This cancellation may not occur when the
parameters of QCD are modified. Of course, other, less extraordinary
causes may be at work: non-zero temperature corrections, or even
algorithmic problems as suggested in~\cite{Azcoiti}. 
This line of investigation should be pursued further.

\section{Acknowledgement}

We thank Olivier Martin for correspondence.
S.K. is supported in part by Korea Science and Engineering Foundation.

\end{document}